\documentstyle{article}

\def\beg{\begin{equation}} \def\ene{\end{equation}}

\begin{document}

\large

$$   $$ \vskip 5cm \centerline{\Large {\bf HEAVISIDE TRANSFORM WITH RESPECT TO
THE MASS IN QCD }} \vskip 2cm \centerline{HIROFUMI YAMADA}

\vskip 1cm \centerline{\it Mathematics department, Chiba Institute of
Technology}
\vskip 12pt \centerline{\it Shibazono 2-1-1, Narashino-shi, Chiba 275} \vskip
12pt \centerline{\it Japan} \vskip 12pt \centerline{\it e-mail:
yamadah@cc.it-chiba.ac.jp} \vskip 2cm

\centerline{\bf Abstract} \vskip 10pt \large \baselineskip 18pt

We propose the use of Heaviside transform with respect to the quark mass to
investigate dynamical aspects of QCD.  We show that at large momentum transfer
the transformed propagator of massive quarks behaves softly and thus the
dominant
effect of explicit chiral symmetry breaking disappears through Heaviside
transform.  This suggests that the massless approximation would be more
convenient to do in the transformed quantity than in the original one.   As an
example of explicit approximation, we estimate the massless value of the quark
condensate.

\newpage 
\baselineskip 22pt 
\noindent       {\bf 1 Introduction}

In this paper we report the result of the application of Heaviside
transform$^{1}$ with respect to the mass to QCD. Let $m$ be the masses of up and
down quarks and the massless value of some quantity, $\Omega(m)$, be of
interest.
 Ordinary approach is to consider the limit that $\lim_{m\rightarrow
0}\Omega(m)$.  However, in the framework of perturbation expansion, the limit
leads to the trivial results and if $m$ is kept non-zero for the
approximation of
the massless value, the effect of explicit chiral symmetry breaking remains and
gives, for example, the hard high energy behavior of the effective quark mass. 
In the present paper, we demonstrate that the use of Heaviside transform with
respect to $m$ solves the above dilemma and provides us a new way of
approximating the massless dynamics.

\vskip 10pt \noindent   {\bf 2 Heaviside transform}

Since our approach is based on Heaviside transformation of perturbative series,
let us state some basic features of the transform.  Heaviside transform of
$\Omega(m)$ is given by the Bromwich integral, \begin{equation} \hat \Omega(\hat
m)=\int^{s+i\infty}_{s-i\infty}{dm \over 2\pi i}{\exp(m/\hat m) \over
m}\Omega(m), \end{equation} where the vertical straight contour should lie
in the
right of all the poles and the cut of $\Omega(m)/m$.  Since the $\Omega(m)/m$ is
analytic in the domain, $Re(m)>s$, $\hat \Omega(\hat m)$ is zero when $\hat
m<0$.
 The Heaviside transform is the inverse of the (second kind of ) Laplace
transform given by \begin{equation} \Omega(m)=m\int^{\infty}_{-\infty}dx\exp(-m
x)\hat\Omega(1/x), \hskip 3mm x={1 \over \hat m}. \end{equation} Since $\hat
\Omega(\hat m)=0$ for $\hat m<0$, the integral (2) reduces to the familiar
one in
which the integration region is $[0,\infty)$.  However the mathematical
manipulation is straightforward for the form (2) because $\hat \Omega(\hat m)$
involves Dirac $\delta$ function in usual cases.

From (2) we find that \begin{equation} \lim_{m\rightarrow
+0}\Omega(m)=\lim_{\hat
m\rightarrow +0} \hat\Omega(\hat m), \end{equation} if the both limits exist.  
Eq.(3) states that the massless value is an invariant of the Heaviside
transform.
 Hence the transformed function can be directly used to approximate $\Omega(0)$
without integration over $\hat m$.  Here suppose that $\Omega(m)$ denotes the
condensate for the quark with its explicit mass $m$.  We note then that, since
the massless limit in the both functions cannot be taken at finite orders of
perturbation expansion due to the bad infra-red behavior of perturbative QCD, we
must choose some non-zero $m$ or $\hat m $ to approximate $\Omega(0)$ or $\hat
\Omega(0)$, respectively.  Then, some difference of the status of approximation
may occur between the two functions.  We will show that, for the approximate
calculation of quark propagator and the quark condensation, the transformed
functions are more convenient than the original ones.  Though $\hat \Omega$ is
not a physical quantity, there is no objection to use it in the approximation of
$\Omega(0)$.

For the sake of later argument let us show transformations of typical
functions. 
Using (1) it is easy to find that \beg m^{-\rho}\stackrel{{\cal
H}}{\rightarrow}{\hat m^{-\rho} \over \Gamma(1+\rho)}\theta(1/\hat m)\hskip 5mm
(\rho>-1), \ene where $\theta(x)$ is the step function, \begin{equation}
\theta(x)=\left\{ \begin{array}{@{\,}ll} 1 & \mbox{$(x>0)$}\\ 0 & \mbox{$(x<0)$}
\end{array} \right. \end{equation} and ${\cal H}$ represents the Heaviside
transformation. By expanding both sides of (4) in powers of $\rho$, we have
\begin{eqnarray} 1 &\stackrel{{\cal H}}{\rightarrow}& \theta(x),\\ \log m
&\stackrel{{\cal H}}{\rightarrow}& -(\gamma+\log x)\theta(x), \end{eqnarray} and
so on where $x=1/\hat m$.  From these results we have the transform of typical
functions appearing in perturbative expansions.  To obtain the formulae, the
following result is useful, \begin{equation} m\Omega(m)\stackrel{{\cal
H}}{\rightarrow} {\partial \hat \Omega(1/x) \over \partial x}.
\end{equation} For
example, using (6) and (8) we find \begin{equation} m^{k}\stackrel{{\cal
H}}{\rightarrow} \delta^{(k-1)}(x),\quad (k=1,2,3,\cdots). \end{equation} From
(7) and (8) we also find \begin{eqnarray} m\log m  &\stackrel{{\cal
H}}{\rightarrow}& -{1 \over x}\theta(x)-(\gamma+\log  x )\delta(x),\\
m^{2}\log m
 &\stackrel{{\cal H}}{\rightarrow}& {1 \over x^{2}}\theta(x)-{2 \over
x}\delta(x)-(\gamma+\log  x )\delta^{'}(x),\\ m^{3}\log m  &\stackrel{{\cal
H}}{\rightarrow}& -{2 \over x^{3}}\theta(x)+{3 \over x^{2}}\delta(x)-{3 \over
x}\delta^{'}(x)-(\gamma+\log  x )\delta^{''}(x). \end{eqnarray} The terms
containing $\delta$ functions play the role of "counter terms" since they cancel
out the divergences coming from the Laplace integration of first terms.

In this paper we confine ourselves with studying the approximate calculation of
QCD quantities at $m=0$.  From now on, we therefore omit the $\delta$ function
terms of the transformed functions since the Laplace integration is un-necessary
in our scheme that $\Omega(0)$ will be approximated by $\hat \Omega(\hat m)$ at
some fixed non-zero $\hat m$.

The result (9) shows that Heaviside transform forces the polynomial of $m$
vanish.  This is desirable because the polynomial of $m$ should not have nothing
to do with the massless physics.  On the other hand the contributions of the
mass-logarithms remains.  Since the mass-log involves $\mu$, the renormalization
parameter, we find that Heaviside transform keeps the effect of ultraviolet
singularity.  This is in accord with the point of view that the ultraviolet
divergence plays the central role in the non-perturbative effects. Strictly
speaking, renormalization-group invariant mass, $\Lambda_{m}$,  corresponding to
$m$ should be used in place of $m$ in (1).  However for notational simplicity we
use $m$ as long as $m$ is proportional to $\Lambda_{m}$.  We may point out that
$\hat m$ obeys the renormalization group equation same as that for $m$.  This is
because the argument of the exponential in (1) and (2) must be renormalization
group invariant (This is obvious if one writes transformations (1) and (2) in
terms of $\Lambda_{m}$ and the corresponding conjugate which is necessarily
renormalization group invariant).

In what follows we first study the high energy behavior of the Heaviside
function
of the effective mass and find that it behaves softly for non-zero $\hat m$ at
large momentum transfer.  This fact suggests an advantage of dealing with the
Heaviside function when one carry out the approximation of the massless case.  
Next, as an example, we perform a rough estimation of the massless value of the
quark condensate.  Through out this paper we use dimensional
regularization$^{2}$
and Landau gauge.

\vskip 10pt \noindent   {\bf 3 High energy behavior of the transformed function
of the effective quark mass}

The inverse of the quark propagator is written as, \beg
S_{F}^{-1}=p-m-\Sigma(m,p)=A(p,m)p-B(p,m)=A(p,m)\Bigl(p-{\cal
M}(p,m)\Bigl), \ene
and the function ${\cal M}(p,m)$ defines the effective quark mass.  Heaviside
transform of the inverse propagator is given as \begin{equation} \hat
S_{F}^{-1}=\hat A(p,\hat m)p-\hat B(p,\hat m), \end{equation} where
\begin{equation} S_{F}^{-1}(p, m)\stackrel{{\cal H}}{\rightarrow}\hat
S_{F}^{-1}(p,\hat m), \hskip 10pt A(p,m)\stackrel{{\cal H}}{\rightarrow}\hat
A(p,\hat m),\hskip 10pt B(p, m)\stackrel{{\cal H}}{\rightarrow}\hat B(p,\hat m).
\end{equation} The transformed function of the effective mass is defined by
\begin{equation} \hat{\cal M}(p,\hat m)={\hat A(p,\hat m) \over \hat B(p,\hat
m)}. \end{equation} It is easy to show that \begin{equation} {\cal M}(p,0)=\hat
{\cal M}(p,0). \end{equation} To clarify the difference between ${\cal M}(p, m)$
and $\hat {\cal M}(p,\hat m)$, let us discuss the high energy behavior of
$\hat{\cal M}(p,\hat m)$ at the one-loop level.

At the one-loop level, the self energy of quarks is given by \begin{equation}
-i(g\mu^{\epsilon})^{2}C_{F}m(D-1){\Gamma(\epsilon) \over
(4\pi)^{D/2}}\int^{1}_{0}dx\Bigl[m^{2}x-p^{2}x(1-x)\Bigl]^{-\epsilon},
\end{equation} where $C_{F}=(N_{c}^{2}-1)/2N_{c}$ and $D=4-2\epsilon$.  Then at
large $-p^{2}$ the inverse of quark propagator behaves as \begin{eqnarray}
S_{F}^{-1}&=&p-{\cal M}(p, m),\\ {\cal M}(p, m)&=&m\Biggl(1-{3C_{F}\alpha \over
4\pi}\Bigl(\log{-p^{2} \over \mu^{2}}-{4 \over 3}\Bigl)\Biggl)+{m^{3} \over
p^{2}} {3C_{F}\alpha \over 4\pi}\Bigl(\log{-p^{2} \over
m^{2}}+1\Bigl)+O\Bigl({m^{4} \over p^{4}}\Bigl),\nonumber \end{eqnarray}
where we
have subtracted the ultraviolet divergence according to ${\overline {MS}}$
scheme$^{3}$ which we use through out this paper.  Heaviside transform of (19)
will be performed by using the following results coming from (6), (9) and (12),
\begin{equation} 1 \stackrel{{\cal H}}{\rightarrow} 1, \hskip 24pt m
\stackrel{{\cal H}}{\rightarrow} 0, \hskip 24pt m^{3}(\log m) \stackrel{{\cal
H}}{\rightarrow} -2\hat m^{3}. \end{equation} We note from (20) that the
so-called hard piece of order $m$ in (19) disappears and the soft term of order
$m^{3}/p^{2}$ survives after the transform.  The reason that the soft term
remains is that it involves $\log(m)$.  The fact that the mass-log exists at the
order $m^{3}/p^{2}$ can be found by differentiating (18) with respect to $m$.
After the differentiation up to three times, one finds the infra-red singularity
when $m\rightarrow 0$.

From (19) and (20) we have \begin{eqnarray} \hat S_{F}^{-1} & = & p-\hat {\cal
M}(p,\hat m),\\ \hat {\cal M}(p,\hat m)&=&{3C_{F}\alpha \over \pi}{{\hat m}^{3}
\over p^{2}}+O(1/p^{4}),\nonumber \end{eqnarray} and find that $\hat {\cal M}$
behaves softly \renewcommand{\thefootnote}{\fnsymbol{footnote}} {\footnote[2]
{\normalsize The soft behavior was also found by the direct calculation in the
generalized Hartree-Fock approach$^{4}$.  Also, we point out that, though  all
sub-leading terms in $1/p^{2}$ are lost at the one-loop level, higher order
contributions would involve $\log m$ and give non-trivial results under ${\cal
H}$ transformation.}}.  We note that, for non zero $\hat m$, $\hat {\cal
M}(p,\hat m)$ simulates the soft behavior of ${\cal M}(p,0)$ with spontaneous
symmetry breaking$^{5}$.  Thus, the dominant effect of explicit chiral symmetry
breaking has been washed out through Heaviside transform.   From the second
equation in (20) the hard piece vanishes to all orders if it is simply
proportional to $m$.  Actually it is pointed out in Ref.6 that, in pole
subtraction schemes, the coefficient functions of operator product
expansion$^{7}$ are analytic in all masses involved.  Then, since the hard piece
belongs to the unit operator and the unit operator cannot involve any mass-log,
the whole part of the order $m$ is just proportional to $m$. Thus, we conclude
that the leading effect of the explicit mass disappears in the Heaviside
function
of the effective mass to all orders. This reveals an advantage of $\hat {\cal
M}(p,\hat m)$ in approximating ${\cal M}(p,0)$, since keeping $\hat m$ non-zero
is compatible with the chiral symmetry.  On the other hand, for ${\cal M}(p,m)$,
we must set $m=0$ to force the hard piece vanish but this makes the perturbative
${\cal M}$ trivial.

To carry out the massless approximation, we must take higher order contributions
into consideration and approximate $\hat {\cal M}(p,0)$ by setting $\hat m$ as
small as possible in $\hat{\cal M}(p,\hat m)$ {\footnote[3] {\normalsize Higher
order contributions would produce powers of $\log \hat m$ at the $\hat
m^{3}/p^{2}$ order term in $\hat{\cal M}$ and the resulting series of $\hat
m^{3}$ times the series of $\log \hat m/\mu$ corresponds to the transformed
quark
condensate.}}. A typical example is given in the next section.

\vskip 10pt \noindent    {\bf 4 Calculation of the quark condensate}

To prove non-perturbative effects, the perturbative series would not contain
enough information.  However, it is not clear whether it is useless in the
approximate calculation or not.  Actually we have found that the  perturbative
series is effective in the approximate calculation of the dynamical mass in the
Gross-Neveu model$^{8}$.  Since the Heaviside transform removes the dominant
effect of the explicit mass which is irrelevant to massless dynamics, we expect
also in QCD that the transformed quantity is more convenient in the approximate
calculation than the original one.

In this section we try to estimate the massless value of the condensate. 
For the
purpose we directly calculate the condensate for the massive case to two-loops
and transform it to evaluate the massless value, assuming the existence of the
non-zero condensate.

As is well known, the naive product, $\sum_{A, \alpha}\bar q^{A, \alpha}(0)
q^{A}_{\alpha}(0)=\bar q q$, is singular ($A$ and $\alpha$ denote color and
Lorentz indices, respectively) and we need to regularize it.  In this paper, we
define the regular product according to ${\overline {MS}}$ scheme$^{9}$.  We
denote thus regularized product as $[\bar q q]$.  In this definition the
two-loop
calculation of the condensate is given by$^{10}$ \begin{equation} \langle
[\bar q
 q] \rangle={N_{c}m^{3}\mu^{-2\epsilon} \over 4\pi^{2}}\Biggl\{1-\log{m^{2}
\over
\mu^{2}}+{3C_{F}g^{2} \over 8\pi^{2}}\biggl(\log^{2}{m^{2} \over \mu^{2}}-{5
\over 3}\log{m^{2} \over \mu^{2}}+constant\biggl)\Biggl\}, \end{equation} where
\begin{equation} [\bar  q  q]=Z\bar q q-{N_{c}m^{3}\mu^{-2\epsilon} \over
4\pi^{2}} \biggl( \hat{1 \over \epsilon}+{C_{F}g^{2} \over
8\pi^{2}}\Bigl(3\hat{1
\over \epsilon}^{2}- \hat{1 \over \epsilon}\Bigl)\biggl), \end{equation} and
\begin{equation} Z=1-{3C_{F}g^{2}(\mu) \over (4\pi)^{2}}\hat {1 \over \epsilon},
\hskip 3mm \hat {1 \over \epsilon}={1 \over \epsilon}-\gamma+ \log(4\pi).
\end{equation} Note that we need non-multiplicative renormalization to make the
condensate finite (see (23)).  As a result, the transformation property of the
product under the renormalization group changes from that of $Z\bar q q$ and
$\langle [\bar q q]\rangle$ no longer satisfies $\mu d \langle [\bar q
q]\rangle/d\mu=-\gamma_{m}\langle [\bar q q]\rangle$ where
$\gamma_{m}m=\mu\partial m/\partial \mu=\Bigl(-{3C_{F} \over
2\pi}\alpha+O(\alpha^{2})\Bigl)m$.

Our task is first making Heaviside transform of (22) and then improving the
result by renormalization group.  First, we point out that, since the
non-multiplicative pole terms are the cube of $m$, they disappear after
Heaviside
transform to give, \begin{equation} \widehat{\langle[\bar  q 
q]\rangle}=\widehat{\langle  Z\bar q q\rangle},  \end{equation} where the hat
denotes the Heaviside transform and we used (see (9)) \begin{equation}
m^{k}\stackrel{{\cal H}}{\rightarrow}0\hskip 4pt (k=positive\hskip 3pt integer).
\end{equation} Since in the pole subtraction scheme the non-multiplicative piece
is proportional to $m^{3}$ and free from the mass-log, eq.(25) holds to all
orders.  Thus, as in the exact massless case, we could use $Z\bar q q$ as the
local product.  As well as the recovery of the soft behavior in the transformed
effective mass, this result shows that the effect of the explicit mass is
reduced
and it would be better to use the transformed quantity to simulate the massless
case.  Now, from (9), (12) and \begin{equation} m^{3}\log^{2}(m) \stackrel{{\cal
H}}{\rightarrow} -2(m/n)^3\biggl( 3+2\Bigl(\log(m/n)-\gamma\Bigl)\biggl),
\end{equation} we find \begin{equation} \langle[\bar  q  q ]\rangle
\stackrel{{\cal H}}{\rightarrow}\widehat{\langle[\bar  q  q
]\rangle}={N_{c}{\hat
m}^{3} \over \pi^{2}} \Biggl[1-{3C_{F}\alpha(\mu) \over \pi}\biggl(\log{\hat
m^{2} \over \mu^{2}}-2\gamma+{13 \over 6}\biggl)\Biggl], \hskip 5mm
\alpha={g^{2}
\over 4\pi}.  \end{equation} We find by the direct operation of $\mu(d/d\mu)$
that the condensate satisfies the renormalization group equation,
\begin{equation} \mu{d \over d\mu}\widehat{\langle[\bar  q  q
]\rangle}=-\gamma_{m} \widehat{\langle[\bar  q  q ]\rangle}, \hskip 3mm
\gamma_{m}=-{3C_{F} \over 2\pi}\alpha+O(\alpha^{2}), \end{equation} up to the
lowest order.   Thus, as implied by (25), the transformed condensate shows the
correct property under the renormalization group.  The solution of (29) is given
by \begin{equation} \widehat{\langle[\bar  q  q ]\rangle}= \chi(\alpha, \hat m,
\{M'\}, \mu)|_{\mu=\mu_{0}}\alpha(\mu)^{-A},\hskip 3mm
\chi=\widehat{\langle[\bar
 q q ]\rangle}_{0}\alpha_{0}^{A}, \hskip 3mm A= {9C_{F} \over 11C_{G}-2n_{F}}
\end{equation} where the subscript $0$ means the value at $\mu=\mu_{0}$ and
$\{M'\}$ denotes the set of explicit masses of other quarks.

Let us improve the large $\hat m_{0}$ behavior of $\chi$ which can be 
settled by
perturbation expansion.  The behavior is improved by adjusting $\mu_{0}$ in
accordance with $\hat m_{0}$.  We impose $\mu_{0}$ to satisfy \begin{equation}
\log{\hat m_{0} \over \mu_{0}}-\gamma+{13 \over 12}=t, \end{equation} where $t$
is a fixed constant which value is yet un-specified. This condition fixes
$\mu_{0}$ as the function of $\hat \Lambda_{m}$ (the renormalization group
invariant mass corresponding to $\hat m$),  $\Lambda$ (finite QCD scale in
${\overline {MS}}$ scheme), and $t$.  Further the $\hat \Lambda_{m}$ dependence
of $\hat m_{0}$ changes from the simple proportional one.  Actually, from the
renormalization group equation and (31), we find the implicit equation for $\hat
m_{0}$, \begin{equation} \hat m_{0}=\hat \Lambda_{m}\alpha(\mu_{0})^{A}=\hat
\Lambda_{m}\Biggl(\beta_{0}\log{\hat m_{0} e^{-\gamma+13/12-t} \over
\Lambda}\Biggl)^{-A}, \end{equation} at this order of expansion.  Since
$\mu_{0}>\Lambda$ to keep the coupling $\alpha_{0}$ positive, $\hat m_{0}$ must
be larger  than $\Lambda e^{+\gamma-13/12+t}(=\hat m^{*})$ from (31).  When
$\hat
\Lambda_{m}$ goes to zero, $\hat m_{0}\rightarrow \hat m^{*}$ and
$\alpha\rightarrow +\infty$ from (31) and (32).  Thus, $\hat m^{*}$ represents
the limitation of the obtained  perturbative result.  Although the
correspondence
between $\hat\Lambda_{m}$ and $\hat m_{0}$ has been involved, $\hat m_{0}$ is a
monotonic function of $\hat \Lambda_{m}$ and most manipulation can be
carried out
in terms of $\hat m_{0}$.  We note from (31) that $\alpha_{0}$ depends on $\hat
m_{0}$ as governed by $\hat m_{0}\partial \alpha_{0}/\partial \hat
m_{0}=\beta(\alpha_{0}) =-\beta_{0}\alpha_{0}^{2}+O(\alpha_{0}^{3})$ where
$\beta_{0}=(11C_{G}-2n_{F})/6\pi$.  The asymptotic freedom in QCD$^{11}$ enters
into the condensate through $\alpha_{0}(\hat m_{0},t)$.  Taking (31) into
account, we thus have \begin{equation} \chi={N_{c}{\hat m}_{0}^{3} \over
\pi^{2}}
\biggl[1-{6C_{F}\alpha_{0}(\hat m_{0}, t) \over \pi}t\biggl]\alpha_{0}(\hat
m_{0}, t)^{A}, \end{equation} and arrive at the improved result,
\begin{equation}
\widehat{\langle[\bar  q  q ]\rangle}= \chi\alpha(\mu)^{-A}={N_{c}{\hat
m}_{0}^{3} \over \pi^{2}} \biggl[1-{6C_{F}\alpha_{0}(\hat m_{0}, t) \over
\pi}t\biggl]\alpha_{0}( \hat m_{0}, t)^{A}\cdot \alpha(\mu)^{-A}. \end{equation}

Let us turn to the rough estimation of the condensate, $\langle[\bar  q  q
]\rangle|_{m=0}$.  Recall that one cannot take the $\hat \Lambda_{m}\rightarrow
0$ limit as we stated in the previous paragraph.  Then what we can do is to fix
$\hat \Lambda_{m}$ or equally $\hat m_{0}$ in terms of $\Lambda$ under the
guiding principle that one should minimize the effect of the explicit mass as
possible as one can.  Here we note that $\lim_{m\rightarrow 0}m^{j}\langle[\bar 
q q ]\rangle=0$ for any posotive integer $j$ and therefore from (3) that
$\lim_{\hat m\rightarrow 0}{\cal H}[m^{j}\langle[\bar  q q ]\rangle]=0$ (${\cal
H}$ denotes the operation of Heaviside transformation).  Let us impose the case
of $j=1$.  Further, since we have another parameter $t$ which we can choose
freely, we demand the condition of $j=2$.  Now from (8), the conditions of $\hat
m_{0}$ minimizing the effect of the explicit mass are \begin{equation} {\partial
\widehat{\langle[\bar  q   q ]\rangle} \over \partial (1/\hat m_{0})}=0, \hskip
3mm {\partial^{2} \widehat{\langle[\bar q q ]\rangle} \over \partial (1/\hat
m_{0})^{2}}=0, \end{equation} which lead to the same conditions for $\chi$.
Eq.(35) gives two non-trivial solutions.  However, one of those corresponds to
the large coupling, $\alpha\sim 8$, and we discard this since it is too large to
rely upon within the two-loop level.  The other solution gives $\alpha\sim 0.8$
for various flavors and is valuable to be discussed further.  That value of
$\alpha$ indicates that the renormalization scale $\mu_{0}$ is a few of
$\Lambda$.  Assuming the decoupling of heavy quarks$^{12}$, we therefore work
with the choice that $n_{F}=3$.  Then we have $\alpha\sim 0.79$ and $t\sim
0.91$.
 These values give $\hat m_{0}\sim 3.6 \Lambda_{3}$, $\chi\sim -10.8
\Lambda_{3}^{3}$ and $\hat \Lambda_{m}\sim 4\Lambda_{3}$ where (31) and the
one-loop relation, $\Lambda_{3}=\mu\exp(-1/\beta_{0}\alpha)$, was used
($\Lambda_{3}$ denotes the QCD scale effective at three flavors).  Thus we
arrive
at \begin{equation} \widehat{\langle[\bar q  q ]\rangle}\sim
-10.8\Lambda_{3}^{3}\alpha^{-A}(\mu). \end{equation}

Further numerical computation of the condensate can be done as follows: From the
experimental data, $\alpha(m_{Z})\sim 0.117\pm 0.005$ $^{13}$, and the
definition
of $\Lambda$ at the lowest order, we have $\Lambda_{5}\sim 83+27-23 MeV$ for
$n_{F}=5$.  With the help of the relation, $\Lambda_{3}\sim
\Lambda_{5}(m_{c}m_{b}/\Lambda_{5}^2)^{2/27}$ ($m_{b}=4.3 GeV,m_{c}=1.3
GeV$)$^{14}$, it is converted to the scale value for $n_{F}=3$, giving
$\Lambda_{3}$ $\sim 135+28-31 MeV$.  By setting $\mu=1 GeV$ and $n_{F}=3$ in
(36), we thus have \begin{equation}  \widehat{\langle[\bar  q q
]\rangle}|_{\mu=1GeV}\sim -(351+88-77 MeV)^{3}. \end{equation}

\vskip 10pt \noindent    {\bf 5 Discussion}

The phenomenological value, $\langle \bar u u\rangle\sim\langle \bar d
d\rangle\sim-(250MeV)^{3}$ $^{15}$ is slightly below the range shown in
(37).  We
think that the result is rather good, taking  into account that the result (37)
is just the lowest order one  which comes from a rough ansatz, (34).  Of course
the higher order calculation is needed for the serious study of our
approach.  In
particular the three loop calculation is important since at this order the
non-linear gluon vertex and the loops of other quarks becomes explicitly active.
\vskip 24pt {\bf Acknowledgements} \baselineskip 20 pt

We wish to thank Dr. O. Morimatsu for the discussion and helpful comments in the
early stage of this work.  We also thanks Dr. H. Suzuki for the interest and
fruitful discussion on the subjects treated in the paper.  We acknowledge the
warm hospitalities at Institute for Nuclear Study, University of Tokyo and
Ochanomizu University.  This work is financially supported by Iwanami Fujukai.

\newpage \baselineskip 20pt \begin{center} {\bf References} \end{center}
\begin{description} \item [{1}] S. Moriguchi et al, Suugaku Koushiki II, Iwanami
Shoten (in Japanese). \item [{2}] G. t'Hooft and M. Veltman, Nucl. Phys. B44
(1972) 189; \\ C. G. Bollini and J. J. Giambiagi, Phys. Lett. B40 (1972) 566;\\
G. M. Cicuta and E. Montaldi, Nuovo Cimento Lett. 4 (1972) 329. \item [{3}]
W. A.
Bardeen, A. J. Buras, D. W. Duke and T. Muta, Phys. Rev. D18 (1978) 3998. \item
[{4}] H. Yamada, Zeit. Phys. C59 (1993) 67. \item [{5}] H. Politzer, Nucl. Phys.
B117 (1976) 397. \item [{6}] K. G. Chetyrkin, S. G. Gorishny and F. V. Tkachov,
Phys. Lett. B119 (1982) 407. \item [{7}] K. Wilson, Phys. Rev. 179 (1969) 1499.
\item [{8}] D.J.Gross and A.Neveu, Phys. Rev. D10 (1974) 3235. \item [{9}] J. C.
Collins, Nucl. Phys. B92 (1975) 477. \item [{10}] V. P. Spiridonov and K. G.
Chetyrkin, Sov. J. Nucl. Phys. 43 (1988) 522. \item [{11}] D. J. Gross and F.
Wilczek, Phys. Rev. Lett. 30 (1973) 1323; \\ H. D. Politzer, Phys. Rev. Lett. 30
(1973) 1346. \item [{12}] T. Appelquist and J. Carazzone, Phys. Rev. D11
(1975)2856. \item [{13}] Review of Particle Properties, Phys. Rev. D50 (1994)
1173. \item [{14}] W. Marciano, Phys. Rev. D29 (1984) 580. \item [{15}] M. A.
Shifman, A. I. Vainshtein and V. I. Zakharov, Nucl. Phys. B147 (1979) 385.
\end{description}

\end{document}